\documentclass[11pt]{article}
\usepackage[dvips]{graphics}
\begin{document}

\noindent
\Large
{\bf RELATIVISTIC QUANTUM MECHANICS AND THE BOHMIAN INTERPRETATION}
\normalsize
\vspace*{1cm}

\noindent
{\bf Hrvoje Nikoli\'c}

\vspace*{0.5cm}
\noindent
{\it
Theoretical Physics Division \\
Rudjer Bo\v{s}kovi\'{c} Institute \\
P.O.B. 180, HR-10002 Zagreb, Croatia \\
E-mail: hrvoje@thphys.irb.hr}

\vspace*{2cm}

\noindent
Conventional relativistic quantum mechanics, based on the Klein-Gordon 
equation, does not possess a natural probabilistic interpretation
in configuration space.
The Bohmian interpretation, in which probabilities play a secondary 
role, provides a viable interpretation of relativistic quantum 
mechanics. We formulate the Bohmian interpretation of many-particle 
wave functions in a Lorentz-covariant way.
In contrast with the nonrelativistic case, the relativistic 
Bohmian interpretation may lead to measurable predictions on particle 
positions even when 
the conventional interpretation does not lead to such predictions. 
\vspace*{0.5cm}


\section{INTRODUCTION}
\label{secI}

It is well known that the probabilistic interpretation 
of the nonrelativistic Schr\"odinger equation for 
particles without spin does not work for its relativistic 
generalization -- the Klein-Gordon equation
(see, e.g., Ref.~\cite{bjor1}).
This is related to the fact that the Klein-Gordon equation
\begin{equation}\label{KG}
(\partial^{\mu}\partial_{\mu}+m^2)\psi(x)=0
\end{equation}
(where $x=(t,\bf{x})$ and we take $\hbar=c=1$) 
contains a second time derivative, instead of the first time 
derivative that appears in the Schr\"odinger equation.
%
%
The quantity $|\psi(x)|^2$ cannot be interpreted as a 
probability density for a particle to have the position $\bf{x}$ 
at time $t$ because then the total probability 
$\int d^3x |\psi(x)|^2$ would not be conserved in time.
One can introduce the conserved current
\begin{equation}\label{cur}
j^{\mu}=i\psi^* \!\stackrel{\leftrightarrow\;}{\partial^{\mu}}\! \psi 
\end{equation}
(where $a \!\stackrel{\leftrightarrow\;}{\partial^{\mu}}\! b \equiv
a\partial^{\mu}b - b\partial^{\mu}a$),
but the time component $j^0(x)$ cannot be interpreted as a probability 
density because it is not positive definite.
Note that this is not a problem for the scattering formalism 
where one assumes that wave functions are positive-frequency 
plane waves asymptotically. However, quantum theory is more 
than a theory of scattering and the problem of negative 
$j^0$ arises outside the scattering regime.  

The usual solution of this problem consists in adopting the 
second quantization of $\psi$ (see, e.g., Ref.~\cite{bjor2}), where 
$\psi$ is not a wave function determining probabilities, but  
an observable (called field) that satisfies the quantum 
uncertainty laws. However, if, at the fundamental level,
$\psi$ should not be interpreted as a wave function 
that determines probabilities of particle positions, then 
it is not clear at all why
such an interpretation of $\psi$ is in such good 
agreement with experiments for nonrelativistic particles.

Using the Bohmian interpretation 
\cite{bohm,bohmPR1,bohmPR2,holPR,holbook} and the theory of 
particle currents \cite{nikcur1,nikcur2,nikcur3},
a theory that consistently combines 
the postulates of second quantization 
(relativistic quantum field theory) with the postulates of 
first quantization 
(quantum mechanics) has recently been proposed in 
Refs.~\cite{nikoldbb1,nikoldbb2}. The comparative advantages of the Bohmian 
interpretation over other interpretations applied to relativistic 
quantum mechanics have been discussed in Ref.~\cite{nikoldbb3}.
In Refs.~\cite{nikoldbb1,nikoldbb2}, the equations that determine 
the Bohmian trajectories of relativistic quantum particles
described by {\em many-particle} wave functions
were written in a form that required a preferred time 
coordinate. Indeed, it is often argued that any relativistic 
hidden variable theory compatible with quantum nonlocality 
must introduce a preferred Lorentz frame. However, as 
demonstrated in Ref.~\cite{bern}, this is not necessarily so.
In the present paper, we further elaborate some of the ideas introduced 
in Refs.~\cite{nikoldbb1} and \cite{bern}
to formulate a fully Lorentz-covariant Bohmian interpretation of 
relativistic quantum mechanics for particles without spin. 
(The generalization to other spins is straightforward.)

As in Refs.~\cite{bern,nikoldbb1}, it appears that particles may 
be superluminal, i.e., faster than light in the vacuum.
Before discussing how this happens in the framework of the relativistic 
Bohmian interpretation, it is important to emphasize that, 
contrary to frequent claims, the principle of Lorentz covariance
does {\em not} forbid superluminal velocities and superluminal 
velocities do {\em not} lead to 
causal paradoxes (see, e.g., Refs.~\cite{lib,nikolcaus}). Indeed, there is  
large evidence that various relativistic interactions 
may cause ordinary particles or waves to propagate superluminally
\cite{gar,chu,drum,sch,chi,bol,chi2,nik}.

As noted in Ref.~\cite{bern}, the Lorentz-covariant Bohmian interpretation 
of the many-particle Klein-Gordon equation is not 
{\em statistically transparent}, i.e., the statistical distribution
of particle positions cannot be calculated in a simple way 
from the wave function alone without the knowledge of  
particle trajectories. This lack of statistical transparency 
is one of the main objects of the present paper, the physical 
meaning of which is qualitatively discussed in Sec.~\ref{secST}.
The quanitative formulation of the Lorentz-covariant Bohmian interpretation
is given in Sec.~\ref{secBM}, while some statistical predictions 
are discussed in Sec.~\ref{secSP}.
The conclusions are drawn in Sec.~\ref{secCon}. 

\section{STATISTICAL TRANSPARENCY}
\label{secST}

Consider a dynamical theory of configuration variables 
that may or may not involve the existence 
of trajectories of configuration
variables. We say that this theory 
is statistically transparent 
if one can calculate the probabilites for possible outcomes 
of measurements of configuration variables in a natural 
way that, in particular, does {\em not} 
refer to any information on the
(possibly existing) trajectories. 
A remarkable property of nonrelativistic 
quantum mechanics (QM) 
is that it {\em is} statistically transparent. 
In other words, one can calculate 
the probability density for particle positions directly from the 
wave function, without knowing particle trajectories. 
Therefore, from the practical calculational point of view, 
the concept of a particle trajectory in nonrelativistic QM 
is simply superfluous. 
This is certainly the main reason 
that the Bohmian interpretation of 
nonrelativistic QM is ignored by most physicists,
and is probably the main reason for
the wide belief that, in nonrelativistic QM, 
particle trajectories simply do not exist.
In other words, statistical transparency is the main reason 
for the wide belief that QM is a fundamentally probabilistic 
theory. Without statistical transparency, there would 
no longer be a good reason for such a belief. 

However, nonrelativistic QM is certainly not the most fundamental 
theory that we know. Is statistical transparency a fundamental 
principle, or just a property of some approximative 
theories? For example, what about relativistic (i.e., Lorentz-covariant)
theories without a preferred time? The following intentionally 
chosen facts suggest that statistical transparency may not be 
a fundamental principle of nature:

\begin{enumerate}
\item Classical mechanics, either nonrelativistic or relativistic, 
is {\em not} statistically transparent.
\item Relativistic quantum mechanics based on the Klein-Gordon 
equation or some of its generalizations is {\em not} 
statistically transparent (owing to the reasons discussed in Sec.~\ref{secI}).
\item The relativistic Bohmian interpretation of the
Dirac equation is statistically transparent, 
but the corresponding
many-particle relativistic generalization is {\em not} 
statistically transparent \cite{bern} (unless a preferred time 
coordinate is determined in a yet unknown dynamical way \cite{durr99}).
\item Nonrelativistic QM is statistically transparent, 
but it is {\em not completely} statistically transparent, 
in the sense that, for a fixed time $t$, it gives the probability  
density $\rho(x^1,x^2,x^3)$, but, for example, 
for a fixed $x^3$, it does not give a probability  
density $\rho(x^1,x^2,t)$.
\item The background-independent quantum gravity based on 
the Wheeler-DeWitt equation lacks the notion of time 
and is {\em not} statistically transparent \cite{kuc,ish}.
\end{enumerate}

Note that, given fact 4, one should not be surprised 
with the fact that relativistic QM may not be statistically tranparent.
Since time and space should play equal roles in a relativistic 
theory, from fact 4 one might expect that 
relativistic QM should be either 
completely statistically transparent or not statistically 
transparent at all. For the Klein-Gordon equation, it is 
the latter possibility that actually realizes.

The lack of statistical transparency does not automatically imply 
that the probabilities cannot be calculated at all. For example, 
in classical mechanics, if we know the probability distribution 
$\rho({\bf x})$ at some initial time and 
the exact particle velocity for each possible initial position 
${\bf x}$, we can calculate $\rho({\bf x})$ at any 
other time by calculating {\em particle trajectories} for all 
possible initial positions. 
(For practical purposes, it is usually sufficient to calculate 
the trajectories for a large but finite sample of initial positions.)
In a similar way, in principle, {\em one can obtain statistical 
predictions from any quantum theory provided that deterministic 
trajectories do exist}. On the other hand, if deterministic 
trajectories do not exist, then it is not clear at all how to assign 
{\em any} physical interpretation to a quantum theory that 
lacks statistical transparency. 
Sinse most of the interpretations 
of QM do not incorporate deterministic trajectories, we conclude 
that, in general, {\em the Bohmian interpretation of quantum theory 
is more powerful than most other interpretations} 
(see also Ref.~\cite{nikoldbb3}).

The lack of statistical transparency in relativistic QM
translates into the lack of statistical transparency in relativistic 
Bohmian mechanics. This feature is often considered as a drawback
of Bohmian mechanics, which motivates investigations of 
various modifications 
of the formalism (e.g., based on the introduction of a preferred 
time coordinate \cite{holpra} or on the representation of the 
Klein-Gordon equation by a first-order differential equation 
\cite{holNC}), 
such that statistical transparency is recovered.
Such modifications typically imply statistical transparency 
even without the Bohmian interpretation, making the Bohmian 
interpretation unnecessary for practical predictions.  

In our view, the lack of statistical transparency  
is not a drawback of a Bohmian theory, 
but rather its {\em virtue}. The reason is the following. 
The statistical predictions of nonrelativistic Bohmian 
mechanics are equivalent to those of the conventional 
interpretation, which makes the scientific value 
of the Bohmian interpretation questionable, because its basic 
assumption - the existence of particle trajectories - cannot 
be verified experimentally. On the other hand, for a quantum 
theory without statistical transparency, the statistical 
predictions of a Bohmian interpretation are {\em not} equivalent 
to those of the conventional interpretation, simply because,
in general,  
the conventional interpretation does not provide any 
statistical predictions for configuration variables at all. 
This opens the possibility
of verifying the validity of a Bohmian interpretation
experimentally,  
i.e., of obtaining some statistical predictions by explicitly 
calculating the trajectories and comparing the predictions with 
experiments. If the predictions turn out to disagree with experiments, 
then one may conclude that this particular Bohmian theory 
is wrong. (One of the main criteria for 
a meaningful scientific theory is that it can be falsified.) 
If the predictions turn out to agree 
with experiments, then one can conclude that the theory 
is correct and that the particle trajectories 
are very likely to really exist (instead of merely being a calculational 
tool \cite{lop}), 
because the same predictions cannot be obtained without 
the trajectories.
     
In Bohmian mechanics without statistical transparency, 
one can obtain statistical predictions at later times if 
the probability distribution $\rho({\bf x})$ is known at the initial 
time. However, how to know this initial $\rho$? Since the theory is not
statistically transparent, in general, there 
is no way of knowing it in a purely theoretical way. 
Fortunately, in some special cases, 
it is possible to know the initial $\rho$ in a purely theoretical way, 
even without statistical 
transparency at all times. For example,
assume that particles 
are slow initially, so that the nonrelativistic approximation
can be used. Then one can invoke either the nonrelativistic 
typicality argument \cite{durr1,durr2} or the nonrelativistic 
quantum H-theorem \cite{val} to conclude that 
the quantum equilibrium takes place initially, i.e., 
that the initial 
probability distribution is given by $\rho=\psi^*\psi$. 
%
%
Then, assume that later interactions are such that particles 
become relativistic, so that $\psi$ is a relativistic solution  
which is not statistically transparent
(i.e., the quantity $j^0$ is not positive definite) 
at later times. In such a case, 
the statistical predictions for particle positions at later times 
can be obtained with the Bohmian interpretation, but not with 
the conventional interpretation.
Such predictions can be compared with experiments.
This motivates us to study the relativistic Bohmian interpretation 
at the quantiative level, which we do in the subsequent sections.

\section{LORENTZ-COVARIANT BOHMIAN MECHANICS}
\label{secBM}

Let $\hat{\phi}(x)$ be a scalar field operator satisfying 
the Klein-Gordon equation (\ref{KG}).
For simplicity, we take $\hat{\phi}$ to be a hermitian uncharged field, 
so that 
the negative values of the time component of the current (to be defined
below) 
cannot be interpreted as negative-charge densities.
Let $|0\rangle$ be the vacuum and $|n\rangle$ an arbitrary 
$n$-particle state. These states are Lorentz-invariant objects.
The corresponding $n$-particle wave function is \cite{schweber,nikoldbb1}
\begin{equation}\label{wf}
\psi(x_1,\ldots ,x_n)=(n!)^{-1/2}S_{\{ x_a\} }
\langle 0|\hat{\phi}(x_1)\cdots\hat{\phi}(x_n)|n\rangle .
\end{equation}
The symbol $S_{\{ x_a\} }$ ($a=1,\ldots ,n$) 
denotes the symmetrization over all $x_a$, 
which is needed because the field operators do not commute 
for nonequal times.
The wave function $\psi$ satisfies $n$ Klein-Gordon equations
\begin{equation}\label{KGn}
(\partial_a^{\mu}\partial_{a\mu}+m^2)\psi(x_1,\ldots ,x_n)=0 ,
\end{equation}
one for each $x_a$.
Although the operator $\hat{\phi}$ is hermitian, the nondiagonal
matrix element $\psi$ defined by (\ref{wf}) is complex.
Therefore,
one can introduce $n$ real 4-currents
\begin{equation}\label{curn}
j^{\mu}_a=i\psi^* \!\stackrel{\leftrightarrow\;}{\partial^{\mu}_a}\! \psi ,
\end{equation}
each of which is separately conserved:
\begin{equation}\label{cons}
\partial^{\mu}_a j_{a\mu}=0.
\end{equation}
Equation (\ref{KGn}) also implies  
\begin{equation}\label{KGs}
\left( \sum_a\partial_a^{\mu}\partial_{a\mu}+nm^2 \right)
\psi(x_1,\ldots ,x_n)=0 ,
\end{equation}
while (\ref{cons}) implies
\begin{equation}\label{conss}
\sum_a\partial^{\mu}_a j_{a\mu}=0.
\end{equation}
Next we write $\psi=Re^{iS}$, where $R$ and $S$ are real
functions. Equation (\ref{KGs}) 
is then equivalent to a set of two real equations
\begin{equation}\label{cont}
\sum_a\partial_a^{\mu}(R^2\partial_{a\mu}S)=0,
\end{equation}
\begin{equation}\label{HJ}
-\frac{\sum_a(\partial_a^{\mu}S)(\partial_{a\mu}S)}{2m} +\frac{nm}{2} +Q=0,
\end{equation}
where
\begin{equation}\label{Q}
Q=\frac{1}{2m}\frac{\sum_a\partial_a^{\mu}\partial_{a\mu}R}{R}
\end{equation}
is the quantum potential. Eq.~(\ref{cont}) is equivalent to 
(\ref{conss}), while (\ref{HJ}) is the quantum analog of the 
relativistic Hamilton-Jacobi equation for $n$ particles. 
The corresponding classical Hamilton-Jacobi equation takes the same
form as (\ref{HJ}), except for the fact that there is no $Q$ term
in the classical case.  
The Bohmian interpretation consists in postulating the existence 
of particle trajectories $x_a^{\mu}(s)$, satisfying
\begin{equation}\label{traj}
\frac{dx_a^{\mu}}{ds} = -\frac{1}{m}\partial_a^{\mu}S .
\end{equation} 
Here $s$ is an affine parameter along the $n$ curves
in the 4-dimensional Minkowski spacetime. These $n$ curves can also 
be viewed as one curve in a $4n$-dimensional configuration 
space. Equation (\ref{traj}) has a form identical 
to that of the corresponding
classical relativistic equation. Equation (\ref{traj}) 
can also be written in another form, as
\begin{equation}\label{traj2}
\frac{dx_a^{\mu}}{ds} = \frac{j_a^{\mu}}{2m\psi^*\psi} .
\end{equation}
From (\ref{traj}), (\ref{HJ}), and the identity 
\begin{equation}
\frac{d}{ds}=\sum_a\frac{dx_a^{\mu}}{ds}\partial_{a\mu},
\end{equation}
one also finds the equation of motion
\begin{equation}\label{eqm}
m\frac{d^2x_a^{\mu}}{ds^2}=\partial_a^{\mu}Q.
\end{equation}

Note that the equations above for the particle trajectories are nonlocal, 
but still Lorentz covariant.
The Lorentz covariance is a consequence of the fact that the trajectories 
in spacetime do not depend on the choice of the affine parameter 
$s$ \cite{bern}. Instead, by choosing $n$ ``initial" spacetime positions
$x_a$, the $n$ trajectories are uniquelly determined by the vector fields 
$j_a^{\mu}$ or $-\partial_a^{\mu}S$. More precisely, 
{\em the trajectories are integral curves of these vector fields}. 
(These two vector fields are equal up to a local positive definite scalar
factor, which implies that they generate the same integral curves.)  
The nonlocality is encoded in the fact that the right-hand 
side of (\ref{eqm}) depends not only on $x_a$, but also 
on all other $x_{a'}$. This is a consequence of the fact that 
$Q(x_1,\ldots ,x_n)$ in (\ref{Q}) is not of the form 
$\sum_a Q_a(x_a)$, which is also closely related to the fact that
$S(x_1,\ldots ,x_n)$ is not of the form 
$\sum_a S_a(x_a)$. The quantities $Q$ and $S$ take these forms 
only when the wave function is a product of the form
$\psi(x_1,\ldots ,x_n)=\psi(x_1)\cdots\psi(x_n)$ (recall that the 
total wave function must be symmetric for bosons). In this case 
there is no entanglement and the nonlocality disappears. 

Note also that the fact
that we parametrize all trajectories with the same parameter $s$ 
is not directly related to the nonlocality, because such a parametrization 
can be used even in local classical physics. Indeed, in a canonical approach 
with a single classical Hamilton-Jacobi equation 
that describes all $4n$ degrees 
of freedom, such a parametrization is the most natural. 
Conversely, 
even in nonlocal quantum physics, once the 
$n$ curves $x^{\mu}_a(s)$ are found, one can reparametrize each 
of the curves with its own parameter $s_a$. However, when the interactions 
are local, then one can even use another parameter for each of the 
$n$ particle {\em equations of motion themselves}. 
For example, one can replace (\ref{traj}) with 
$dx_a^{\mu}/ds_a = -\partial_a^{\mu}S_a/m$. On the other hand, 
when the interactions 
are not local, then one must use a single parameter $s$ in the 
equations of motion for the trajectories, whereas new separate parameters 
$s_a$ can be used only after these equations of motion have been solved.

Now consider the nonrelativistic limit. In this limit, {\em all}  
wave-function frequencies are (approximately) equal to $m$, so 
from (\ref{curn}) one finds that {\em all}
time components of the currents are equal and given by 
$j^0_a=2m\psi^*\psi\equiv\tilde{\rho}$, which does not depend on $a$. 
Therefore, 
in the nonrelativistic limit, the time components of the currents 
become positive definite and unique for all particles. 
By introducing the quantity
\begin{equation}
\rho({\bf x}_1,\ldots ,{\bf x}_n,t)=
\tilde{\rho}(x_1,\ldots ,x_n)|_{t_1=\cdots =t_n=t}, 
\end{equation}
one finds that the nonrelativistic limit of (\ref{conss}) can be written 
as    
\begin{equation}\label{conssnr}
\frac{\partial\rho}{\partial t}+  \sum_a\partial^{i}_a j_{ai}=0,
\end{equation}   
where $i=1,2,3$ label space coordinates. Equation (\ref{conssnr}) 
implies that $\rho$ can be interpreted as the probability density, 
which explains why the nonrelativistic limit leads to  
statistical transparency, but not to complete statistical 
transparency. 

In general, 
in the full relativistic case, there is no analog of the function 
$\rho$ that would be both positive definite and independent of $a$. 
Such a quantity exists only in some special cases.
For example, the independence of 
$j^0_a$ on $a$ occurs when the wave function is a product 
$\psi(x_1,\ldots ,x_n)=\psi(x_1)\cdots\psi(x_n)$, while the 
positivity of $j^0_a$ occurs when $\psi(x)$ is a plane wave 
(not a superposition of plane waves) $\psi(x)=e^{-ip_{\mu}x^{\mu}}$
with a positive frequency $p_0$. 
%
%
In such special cases, 
one can introduce the quantity $\rho$ in a way similar to that 
of the nonrelativistic limit, leading to an equation 
of the form (\ref{conssnr}) valid also in the relativistic 
case.

Note also that the 4-currents $j^{\mu}_a$ do not need 
to be timelike, but locally can be lightlike or even spacelike.
This implies that massive particles may move with the 
velocity of light or even faster. As shown in Ref.~\cite{nikoldbb1},
such superluminal velocities cannot be observed, so there is no 
contradiction with experiments.  
(To understand why these superluminal velocities cannot be observed, 
one has to take into account the theory 
of quantum measurements. To emphasize the role of measurements, 
we also note that, contrary to frequent claims,
the theory of quantum measurements 
{\em is} crucial for understanding why nonrelativistic Bohmian 
mechanics is consistent not only
with the conventional statistical predictions for particle positions,
but also with {\em all} statistical 
predictions of the conventional interpretation
\cite{bohm,bohmPR1,holbook,nikoldbb1}.)  

\section{TOWARDS STATISTICAL PREDICTIONS}
\label{secSP}

In order to understand more closely how interesting 
statistical predictions may be obtained from relativistic
Bohmian mechanics, we study the one-particle case described by a 
wave function $\psi(x)$. The corresponding
current is $j^{\mu}(x)$, for which we assume regularity  
at all $x$. In this case,
one can make statitistical predictions without calculating 
all trajectories, provided that certain additional 
assumptions are fulfilled. 

\begin{figure}[t]
\centerline{\includegraphics{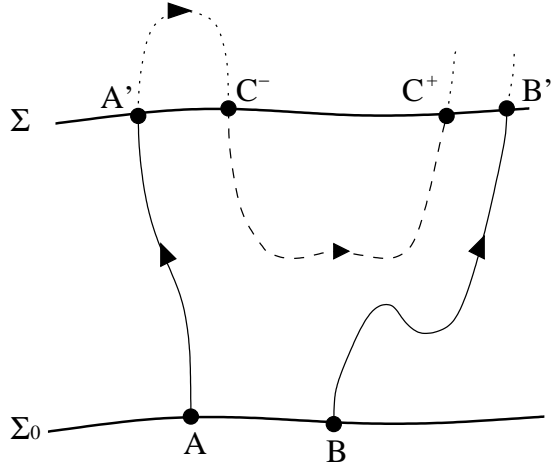}}
\caption{A sample of typical relativistic Bohmian trajectories.
The solid ones represent physically realizable trajectories. 
The dotted ones are unphysical because the measurement takes place 
at the hypersuface $\Sigma$ and above. The dashed one is unphysical 
because it is assumed that only one particle exists and that 
this particle crosses the initial hypersurface $\Sigma_0$.
The arrows indicate the direction of $j^{\mu}$.}
\label{fig1}
\end{figure}

Let $n_{\mu}$ be the unit 
future-oriented timelike vector normal to 
an initial 3-dimensional spacelike
hypersurface $\Sigma_0$. Assume that the quantity 
\begin{equation}\label{j}
j=j^{\mu}n_{\mu}
\end{equation}
has the property $j\geq 0$
everywhere on $\Sigma_0$. 
%
%
Assume also                                                 
that the statistical distribution of particle positions on
$\Sigma_0$ is given by $j$.
We take the normalization such that
\begin{equation}\label{globcons}
\int_{\Sigma_0} dS_{\mu}j^{\mu}=
\int_{\Sigma_0} d^3x \sqrt{|g^{(3)}|}\, j =1,
\end{equation}
where $dS_{\mu}=d^3x \sqrt{|g^{(3)}|}n_{\mu}$ 
is the covariant measure of the volume 
on $\Sigma_0$ and $g^{(3)}$ is the determinant of the induced metric 
on $\Sigma_0$. 
The normalization above corresponds
to the assumption that we know with certainty that 
there is one and only one pointlike particle on $\Sigma_0$.
Let the measurement of particle 
positions be performed at a later spacelike hypersurface $\Sigma$.
Given that the initial statistical distribution is given by
$j$ on $\Sigma_0$, 
what can we conclude about the statistical distribution of particle
positions on $\Sigma$? From the local conservation 
$\partial_{\mu}j^{\mu}=0$, one concludes that (\ref{globcons}) 
is valid also on $\Sigma$:
\begin{equation}\label{globcons1}
\int_{\Sigma} dS_{\mu}j^{\mu}=  
\int_{\Sigma} d^3x \sqrt{|g^{(3)}|}\, j =1,
\end{equation}
where $j$ is defined by (\ref{j}) with respect to an analogous 
normal $n_{\mu}$ on $\Sigma$.
If $j\geq 0$ on $\Sigma$, then 
the statistical distribution on $\Sigma$ will be given by $j$. 
In this case, we have statistical transparency on $\Sigma$.
However, a more interesting question is what if 
$j<0$ on some regions of $\Sigma$?
From (\ref{globcons1})
it is clear that the inequality $j<0$ cannot be valid 
everywhere on $\Sigma$. Let $\Sigma^-$ be the set of all points 
on $\Sigma$ at which $j<0$.
For every point on $\Sigma^-$ there exists a unique point on 
$\Sigma-\Sigma^-$, such that these two points are connected with a 
trajectory in ${\cal M}$ (where ${\cal M}$ is the region of  
Minkowski spacetime bounded by $\Sigma_0$ and $\Sigma$). 
Let $\Sigma^+$ be the set of all such points 
on $\Sigma-\Sigma^-$ that are connected with a point on $\Sigma^-$. 
Finally, let 
\begin{equation}
\Sigma'=\Sigma -(\Sigma^+\cup\Sigma^-) .
\end{equation}
All this is illustrated in Fig.~\ref{fig1}. 
The point $C^-$ is an element of $\Sigma^-$, the point
$C^+$ is an element of $\Sigma^+$, while the points
$A'$ and $B'$ are elements of $\Sigma'$.
The system is described by the wave function $\psi(x)$ 
on ${\cal M}$, 
i.e., between $\Sigma_0$ and $\Sigma$.
The arrows on the 
integral curves in ${\cal M}$ indicate the direction of $j^{\mu}$. 
The dotted curves above $\Sigma$ indicate the particle trajectories 
that might be realized if there were no measurement of particle 
positions on $\Sigma$, i.e., if the system were described by $\psi(x)$ 
even above 
$\Sigma$. If there were no measurement on $\Sigma$, and if the 
initial position of the particle were the point $A$ on Fig.~\ref{fig1}, 
then 
the particle would cross $\Sigma$ at 3 points, i.e., at $A'$, $C^-$, and 
$C^+$. However, owing to the measurement of particle positions, 
the actual wave function above $\Sigma$ is of the form $\psi(x,y)$, 
where $y$ represents the degrees of freedom of the measuring apparatus.
The interactions with the measuring apparatus 
are such that the particles enter localized channels
\cite{bohm,bohmPR1,holbook,nikoldbb1} which typically forbid 
trajectories such as the trajectory connecting $A'$ with $C^-$. This is how 
the theory of quantum measurements explains why the 
Bohmian motions backwards 
in time are not in contradiction with the fact that we do not observe 
multiple copies of particles, such as $A'$, $C^-$, and $C^+$ (see also
\cite{nikoldbb1}). Therefore, in the 
rest of the discussion we ignore the dotted trajectories.
Since we assume 
that the initial distribution is given by $j$ on $\Sigma_0$
(recall (\ref{globcons}) and its interpretation),
the dashed trajectory connecting $C^-$ with $C^+$ 
cannot be realized as an actual particle trajectory either. Only 
the solid trajectories can be realized as the actual trajectories.  

In order to find the probability distribution of particle positions 
on $\Sigma$, first recall that $\Sigma$ can be decomposed in 
3 disjunct sets as
\begin{equation}
\Sigma=\Sigma' \cup \Sigma^+ \cup \Sigma^- .
\end{equation}
Thus the integral (\ref{globcons1}) has contributions from 3 regions. 
These contributions have the properties
\begin{equation}\label{globcons2}
 \displaystyle\int_{\Sigma'} dS_{\mu}j^{\mu}=1, 
\end{equation}
\begin{equation}
 \displaystyle\int_{\Sigma^+} dS_{\mu}j^{\mu} + 
\displaystyle\int_{\Sigma^-} dS_{\mu}j^{\mu}=0. 
\end{equation}
Second, note that a trajectory in ${\cal M}$ crosses $\Sigma'$ if and only 
if it crosses $\Sigma_0$. (See Fig.~\ref{fig1}, recall the definition 
of $\Sigma'$ and the fact that  
$j\geq 0$ on $\Sigma_0$, and note that the integral 
curves of $j^{\mu}$ define a nonsingular congruence on ${\cal M}$.)  
Having this in mind and recalling (\ref{globcons2}) and the 
local conservation law $\partial_{\mu}j^{\mu}(x)=0$ valid on ${\cal M}$, 
it becomes evident
that the probability distribution $\rho({\bf x})$ on $\Sigma$ is given by
\begin{equation}\label{probdis}
\rho=\left\{
\begin{array}{ll}
j & \mbox{on $\Sigma'$}, \\
0 & \mbox{on $\Sigma^+\cup\Sigma^-$}.
\end{array}  
\right.
\end{equation}
This is the {\em measurable} probability distribution.
The remarkable result emerging from the existence of particle trajectories
is that the probabilty for a particle to be found on the region  
$\Sigma^+\cup\Sigma^-$ vanishes, despite the fact that $j$ does not 
vanish there. This is valid even if the initial distribution on 
$\Sigma_0$ is not given by $j$, only provided that we know 
with certainty that the particle has some 
(unknown) position on $\Sigma_0$. 
Then the particle cannot be found on $\Sigma^+\cup\Sigma^-$ simply because 
there is no trajectory in ${\cal M}$ connecting $\Sigma_0$ with 
$\Sigma^+\cup\Sigma^-$.  
The prediction that the particle cannot be found on $\Sigma^+\cup\Sigma^-$
could be tested experimentally and the experimental confirmation
of such a prediction would be a strong support 
for the claim that particles 
really do have trajectories given by the integral curves of $j^{\mu}$.
    
It is also interesting to note that in order to find the distribution 
(\ref{probdis}), the only thing that cannot be found without 
the explicit calculation of the trajectories is the region $\Sigma^+$.
Therefore, one does not need to calculate the physically 
realizable trajectories (the solid ones in Fig.~\ref{fig1}), but only 
the unphysical ones (a dashed one in Fig.~\ref{fig1}) connecting 
points in $\Sigma^-$ with points in $\Sigma^+$. 

Of course, in order to give a more concrete proposal for an experiment
that could verify the validity of the relativistic Bohmian interpretation,
one has to make a more detailed and realistic
quantitative analysis of some specific case (probably with spin) 
in which
the predictions of the Bohmian interpretation are sufficiently different
from those of some other interpretation. Such a specific
analysis is beyond the scope of this paper, but
we hope that the presented ideas will motivate a more extensive research
towards a possible experimental verification of the relativistic
Bohmian mechanics.

\section{CONCLUSION}
\label{secCon}

The Klein-Gordon equation is not statistically transparent, 
i.e., it is not clear how to calculate the probabilities 
of particle positions from the knowledge of the wave function.
The Bohmian interpretation, in which the probabilites only play 
a secondary role, provides a viable interpretation 
of the Klein-Gordon equation. It turns out that the 
lack of statistical 
tranparency is not a drawback but rather a virtue of 
the Bohmian interpretation, because it opens the possibility 
of experimentally distinguishing its predictions 
from the predictions of other possible interpretations. 
Another remarkable result is that 
the equations for Bohmian particle trajectories 
are nonlocal, but they can still be naturally  
written in a Lorentz-covariant 
form without a preferred Lorentz frame.  

\vspace{0.4cm}
\noindent
{\bf Acknowledgements.}
This work was supported by the Ministry of Science and Technology of the
Republic of Croatia under Contract No.~0098002.


\begin{thebibliography}{99}

\bibitem{bjor1}
J.~D.~Bjorken and S.~D.~Drell, {\it Relativistic Quantum Mechanics}
(McGraw-Hill, New York, 1964).
\bibitem{bjor2}
J.~D.~Bjorken and S.~D.~Drell, {\it Relativistic Quantum Fields}
(McGraw-Hill, New York, 1965).
\bibitem{bohm}
D.~Bohm, {\it Phys.~Rev.}~{\bf 85}, 166, 180 (1952).
\bibitem{bohmPR1}
D.~Bohm and B.~J.~Hiley, 
{\it Phys.~Rep.}~{\bf 144}, 323 (1987).
\bibitem{bohmPR2}
D.~Bohm, B.~J.~Hiley, and P.~N.~Kaloyerou, 
{\it Phys.~Rep.}~{\bf 144}, 349 (1987).
\bibitem{holPR}
P.~R.~Holland, {\it Phys.~Rep.}~{\bf 224}, 95 (1993).
\bibitem{holbook}
P.~R.~Holland, {\it The Quantum Theory of Motion}
(Cambridge University Press, Cambridge, 1993).
\bibitem{nikcur1}
H.~Nikoli\'c, {\it Phys.~Lett.}~B {\bf 527}, 119 (2002);
Erratum {\bf 529}, 265 (2002).
\bibitem{nikcur2}
H.~Nikoli\'c, {\it Int.~J.~Mod.~Phys.}~D {\bf 12}, 407 (2003).
\bibitem{nikcur3}
H.~Nikoli\'c, hep-th/0205022, to appear in 
{\it Gen.~Rel.~Grav.}
\bibitem{nikoldbb1}
H.~Nikoli\'c, {\it Found.~Phys.~Lett.}~{\bf 17}, 363 (2004).
\bibitem{nikoldbb2}
H.~Nikoli\'c, quant-ph/0302152, 
to appear in {\it Found.~Phys.~Lett.}
\bibitem{nikoldbb3}
H.~Nikoli\'c, quant-ph/0307179.
\bibitem{bern}
K.~Berndl, D.~D\"urr, S.~Goldstein, and N.~Zangh\`i, 
{\it Phys.~Rev.}~A {\bf 53}, 2062 (1996).
\bibitem{lib}
S.~Liberati, S.~Sonego, and M.~Visser, {\it Ann.~Phys.}~{\bf 298}, 
167 (2002).
\bibitem{nikolcaus}
H.~Nikoli\'c, gr-qc/0403121.
\bibitem{gar}
C.~G.~B.~Garrett and D.~E.~McCumber, 1970 {\it Phys.~Rev.}~A {\bf 1}, 
305 (1970).
\bibitem{chu}
S.~Chu and S.~Wong, {\it Phys.~Rev.~Lett.}~{\bf 48}, 738 (1982).
\bibitem{drum}
I.~T.~Drummond and S.~J.~Hathrell, {\it Phys.~Rev.}~D {\bf 22}, 
343 (1980).
\bibitem{sch}
K.~Scharnhorst, {\it Phys.~Lett.}~B {\bf 22}, 354 (1990).
\bibitem{chi}
T.~Y.~Chiao, {\it Phys.~Rev.}~A {\bf 48}, R34 (1993).
\bibitem{bol}
E.~Bolda, R.~Y.~Chiao, and J.~C.~Garrison, {\it Phys.~Rev.}~A {\bf 48}, 
3890 (1993).
\bibitem{chi2}
R.~Y.~Chiao, A.~E.~Kozhekin, and G.~Kurizki,  
{\it Phys.~Rev.~Lett.}~{\bf 77}, 1254 (1996).
\bibitem{nik}
N.~Bili\'c and H.~Nikoli\'c, {\it Phys.~Rev.}~D {\bf 68}, 085008
(2003).
\bibitem{durr99}
D.~D\"urr, S.~Goldstein, K.~M\"unch-Berndl, and N.~Zangh\`i, 
{\it Phys.~Rev.}~A {\bf 60}, 2729 (1999).
\bibitem{kuc}
K.~Kucha\v r, in: {\it Proceedings of the 4th Canadian Conference 
on General Relativity and Relativistic Astrophysics} 
(World Scientific, Singapore, 1992).
\bibitem{ish}
C.~J.~Isham, gr-qc/9210011.
\bibitem{holpra}
P.~Holland, {\it Phys.~Rev.}~A {\bf 60}, 4326 (1999).
\bibitem{holNC}
P.~R.~Holland and J.~P.~Vigier, {\it Nuovo Cimento} {\bf 88B}, 
20 (1985).
\bibitem{lop}
C.~L.~Lopreore and R.~E.~Wyatt, {\it Phys.~Rev.~Lett.}~{\bf 82}, 
5190 (1999).
\bibitem{durr1}
D.~D\"urr, S.~Goldstein, and N.~Zangh\`i,
{\it J.~Stat.~Phys.}~{\bf 67}, 843 (1992).
\bibitem{durr2}
D.~D\"urr, S.~Goldstein, and N.~Zangh\`i, 
{\it Phys.~Lett.}~A {\bf 172}, 6 (1992).
\bibitem{val} 
A.~Valentini, {\it Phys.~Lett.}~{\bf A} 156, 5 (1991).
\bibitem{schweber}
S.~S.~Schweber, {\it An Introduction to Relativistic Quantum Field Theory}
(Harper \& Row, New York, 1961).

\end{thebibliography}
\end{document}